\documentclass[onecolumn]{aa}
\usepackage{graphicx}
\usepackage[authoryear]{natbib}
\bibpunct[]{(}{)}{;}{a}{}{,}
\usepackage{txfonts}
\newcommand{\mic}{\mbox{$\mu$m}}
\newcommand{\sbs}{\mbox{SBS\,0335-052}}
\newcommand{\ks}{\mbox{Ks}}
\begin{document}

   \title{Dust and Super Star Clusters in NGC~5253
          \thanks{Based on observations obtained at the ESO
           telescopes of La Silla and Paranal, program 69.B-0345; and
           on observations with ISO, an ESA project with instruments
           funded by ESA Member States (especially the PI countries:
           France, Germany, the Netherlands and the United Kingdom)
           and with the participation of ISAS and NASA.}
         }

   \subtitle{}

   \author{L. Vanzi
          \inst{1}
          \and
           M. Sauvage
          \inst{2}
          }

   \offprints{L. Vanzi}

   \institute{ESO - European Southern Observatory
              Alonso de Cordova 3107, Santiago - Chile\\
              \email{lvanzi@eso.org}
         \and
             Service d'Astrophysique, CEA/DSM/DAPNIA, Centre d'Etudes
             de Saclay,
             F-91191 Gif-sur-Yvette Cedex, France\\
             \email{msauvage@cea.fr}
             }

   \date{Received July 5, 2003; accepted }

   \abstract{

   We present new observations of the famous starburst galaxy NGC~5253
   which owes its celebrity to possibly being the youngest and closest
   starburst galaxy known. Our observations in the infrared and
   millimeter contribute to shed light on the properties of this
   interesting object. We have used our new data along with data from
   the literature to study the properties of the young stellar
   clusters present in NGC~5253. We find that the brightest optical
   clusters are all characterized by a near-infrared excess that is
   explained by the combined effect of extinction and emission by
   dust. For the brightest infrared cluster we model the spectral
   energy distribution from the optical to the radio. We find that
   this cluster dominates the galaxy emission longward of 3\,\mic, that it
   has a bolometric luminosity of $1.2~10^9~L_{\odot}$
   and a mass of $1.2~10^6~M_{\odot}$, giving $L/M\approx10^3$. The cluster
   is obscured by 7 mag of optical extinction produced by about
   1.5\,10$^{5}$ $M_{\odot}$ of dust. The dust properties are peculiar
   with respect to the dust properties in the solar neighbourhood
   with a composition
   characterized by a lack of silicates and a flatter size distribution
   than the standard one, i.e. a bias toward larger grains. We find
   that NGC\,5253 is a striking example of a galaxy where the
   infrared-submillimeter and ultraviolet-optical emissions originate
   in totally decoupled regions of vastly different physical sizes.

   }

   \maketitle
%

\section{Introduction}

It has been found that very young clusters can be deeply embedded in dusty
cocoons even when hosted in extremely low metallicity environments. The most
prominent example of this class of objects is certainly the Super-Star Cluster
(SSC) observed in the extremely metal deficient blue compact galaxy
SBS~0335-052 \citep{thu99,hun01}. Modeling the SED of this object indicates
that the radiation from a few \mic\ to the far-infrared (FIR) comes from a
luminous ($>10^{9}$\,L$_{\odot}$), massive ($>10^{6}$\,M$_{\odot}$) star
cluster embedded in a dust cocoon where large grains dominate, contrary to
interstellar medium dust \citep{pla02}. The amount of dust derived by fitting
the SED, about $10^5~M_{\odot}$, is surprisingly high for a very low
metallicity environment and it opens several interesting questions about the
star formation history of the galaxy and the processes involved in dust
formation. Clusters with similar properties and possibly even more embedded,
since they are not detected at $\sim$2\,\mic, have been observed in the blue
compact galaxy He~2-10 \citep{vac02}. It is found that these clusters have
properties very similar to those of galactic ultra dense HII regions and that
they are probably the main source of the observed far-infrared flux. In
SBS~0335-052, the dust-enshrouded SSC contains about 10 times more O7V stars
than its visible counterparts \citep{pla02}. In He~2-10, the embedded source
represent $>1/9^{\rm th}$ of the total UV-emitting star content of the galaxy
\citep{kbj99,con94}. In these two galaxies, the visible super-star clusters
account for almost all of the recent visible star formation activity
\citep{con94,thu97}. Therefore these two examples show that a substantial, if
not dominant, fraction of the current star formation can be made invisible in
the optical-UV by dust obscuration even in low-metallicity environments.
Were this to be a general property, it would obviously have important 
consequences on the interpretation of the high redshift universe where most
star formation should occur in low-metallicity environments.

In this paper, we study the case of NGC\,5253, with the purpose of
demonstrating that it is also a member of the class of galaxies that
host dust-enshrouded SSCs and to strengthen the relevance of the
phenomenon in the study of starburst objects.

The starburst galaxy NGC~5253 is a blue dwarf irregular galaxy which
has gained increasing interest during the years because of a few
important facts.  (1) Among the starburst galaxies it is one of the
closest examples. Its exact distance is still a matter of debate, with
the most recent values being either 3.3$\pm$0.3\,Mpc \citep{gib00} or
4.0$\pm$0.3\,Mpc \citep{thi03}. Here we arbitrarily choose to use a
distance of 3.3\,Mpc, for which 1\arcsec is 16\,pc. (2) It is most likely
the youngest starburst known with a large number of clusters detected
whose ages are in the 2-50\,Myr range \citep[see
e.g.][]{van80,gor96,cal97}. (3) Although its metal content is not
extremely low, as often observed in dwarf galaxies, its metallicity is
sub-solar, about 1/6 $Z_{\odot}$ \citep{kob99}. The starburst nature
of this galaxy finds a direct confirmation in the supernovae
observed. Two SNe have exploded in less than a century, in 1895 and in
1972, which is a fairly high number given that NGC\,5253 is a small
galaxy: its estimated mass is 6\,10$^{9}$\,M$_{\odot}$ \citep[][,
scaled to the adopted distance]{wel70}.
 
As NGC~5253 has been observed by many authors over a wide spectral range
for many decades, a precise picture of the galaxy can be built today
\citep[see e.g.][]{cal89,mar95,cal97,cal99}. We will highlight here
only those observations that are relevant to our present work.

\citet{rie72} observed NGC~5253 at 10 and 20\,\mic\ as part of their
seminal work on infrared photometry of galaxies. \citet{gla73} and
\citet{moo82} also obtained near and mid-infrared observations and
attempted to constrain the spectral energy distribution (SED) in the
optical and infrared. \citet{roc91}, based on their mid infrared spectrum
which shows strong [SIV] but undetected [NeII], classified NGC~5253,
together with II~Zw~40, as a high excitation HII galaxy characterized
by a young starburst and by the presence of very massive stars. A more
complete mid-IR spectrum was published by \citet{cro99} with ISOSWS,
and shows that the spectrum is dominated by [SIV], [NeIII] and [SIII],
a common feature of high excitation low-metallicity HII regions
observed with ISO \citep[e.g.][ in the SMC]{con00}.  Finally, NGC~5253
is detected in all four IRAS bands \citep{mos90}.

A large number of bright clusters was observed by \citet{van80} 
and interpreted as the remnant of a powerful episode of star formation. 
More recently high spatial resolution observations have allowed to
focus the attention on a few peculiarities. \citet[][, hereafter
C97]{cal97} studied the properties of the brightest optical clusters
detected in the HST images. They use broad and narrow band colors and
find that all clusters are young with ages from few to a few tens of
Myr. The youngest cluster is obscured by about 9 mag in V and in
general there is anti-correlation between the presence of dust, that
is observed all over the central part of the galaxy, and the age of
the clusters, i.e. the younger the clusters, the higher the
extinction. The centimetric radio spectrum is entirely thermal
\citep{bec96}, which is highly unusual for a galaxy, and \citet{tur00}
detect a very compact radio source, 1-2 pc in size, which is very
bright at 1.3-2 cm. They attribute this source to a nebula ionized by
about 4000 O7 stars. The ionizing flux derived from the radio is about
two orders of magnitude larger than indicated by the H$\alpha$
emission. \cite{gor01} report their Keck telescope detection of a
bright infrared source at 11.7 and 18.7\,\mic, which they identify
with the compact radio nebula.

The paper is organized as follows: in Sect. \ref{sec:obs} we
describe the new observations on which this work is based. In
Sect. \ref{sec:col} we discuss the optical colors of the
individual bright clusters in the galaxy and exemplify the impact of
dust on these colors. In Sect. \ref{sec:sed} we build and model the
spectral energy distribution of the brightest source in the galaxy and
show that it can be represented by a dust-enshrouded SSC model. The
conclusions of our work are summarized in Sect. \ref{sec:conc}.

\section{Observations}
\label{sec:obs}

We have observed NGC~5253 with ISAAC at the ESO-VLT in the \ks\ and L'
near-infrared (NIR) broad bands on April 19, 2002.  ISAAC is equipped
with a Hawaii $1024\times1024$ Rockwell detector for the short wavelengths
(1-2.5\,\mic) and with an Aladdin SBRC $1024\times1024$ detector for the long
wavelengths (2.5-5\,\mic), the scale is 0.148 and 0.071 \arcsec/pix
respectively. Both observations were obtained under seeing better than
0.4\arcsec. The total on source integration time was 5 minutes in \ks\ and
about 30 minutes in L'. We have used the usual nodding technique to subtract
the background in the \ks\ band and the nod and chop in L'. The images have
been reduced both with the ESO package Eclipse and with IRAF, the two
independent techniques gave perfectly consistent results. The photometric
calibration relied on the observation of one photometric standard star. Since
the Aladdin detector is affected by a significant non linearity, a correction
for this effect has been applied to the L' band observations. In both filters
we have reached a photometric accuracy better than 3\% where most of the error
is due to the non-uniformity across the field. We have checked the
consistency of our photometry and calibrations using the data available in the
literature. \citet{moo82} and \citet{for92} provide in their work aperture
photometry of NGC~5253 on apertures of 3, 6, 9, 12, 36 and 48\arcsec diameter
in Ks and 7.5, 9 and 12\arcsec  diameter in L. Using the same apertures on our 
calibrated images, we find values within less than 0.01 mag of  \citet{for92}
and 0.02 mag on average of \citet{moo82} in K. In L, we can only compare with
\cite{moo82}, the average difference between the two datasets is less than 0.03
mag even  though large individual variations are present that can be 
attributed to centering differences and to the large error bars of the 
measures of \citet{moo82}.

\begin{figure*}
\centering
\includegraphics[width=\textwidth]{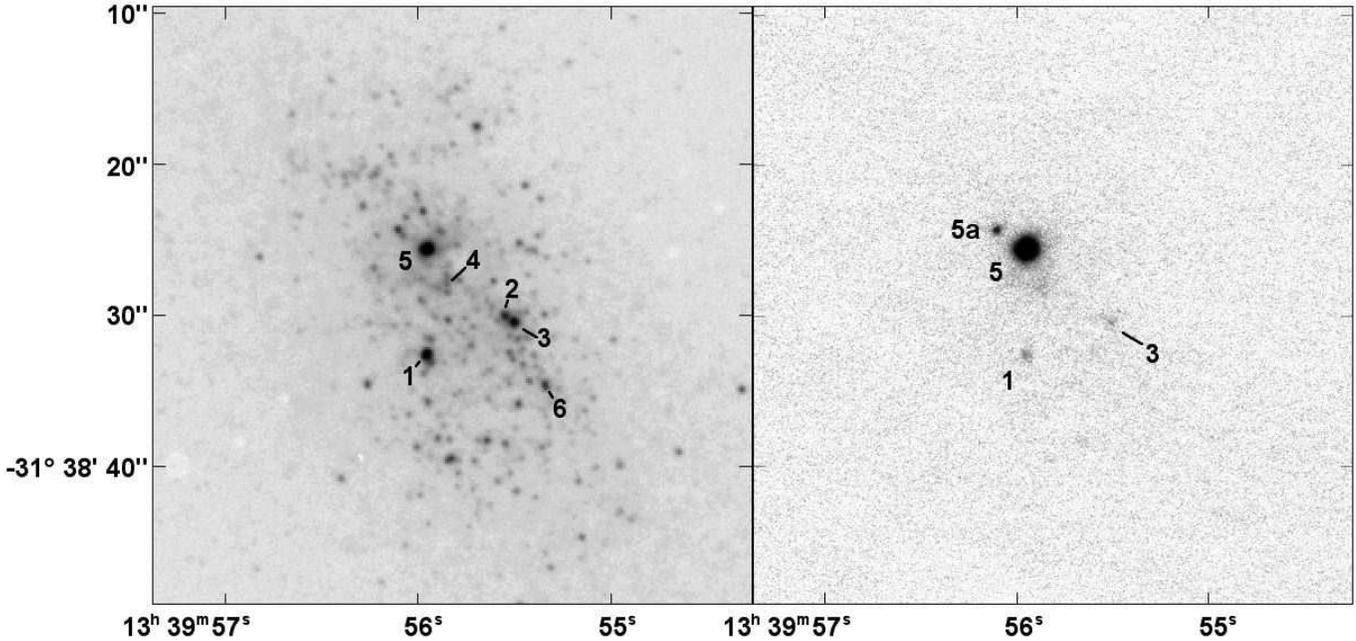}

\caption{NGC~5253 in \ks\ (left) and L' (right) bands. North is up and East is
left. Coordinates refer to the J2000 equinox. The field of view of the image is
40\arcsec. The intensity scale on the \ks\ band has been chosen to clearly show
the brightest clusters. The actual image is much deeper than this display. 
The clusters are labeled as in C97. As the numbers in C97
correspond to a brightness scale, these images clearly show the dramatic
changes occurring in the spectral energy distribution of most clusters in
NGC\,5253. Cluster\,5a is the relatively bright object in the L' image
North-East of Cluster\,5. Note that the intensity levels have been adjusted in
the L' image to show clusters 1 and 3, but that cluster 5 is a point source in
L' as well. Wavelet filtering of the L band image reveals that cluster 4
is responsible for the extension seen SW of cluster 5, though the statistical
significance of this detection is too low to allow a reliable photometric
measurement. At an even lower significance level, the same technique indicates
the possible dection of a source around $\alpha=13^{\rm h}39^{\rm m}55.6^{\rm
s}$ and $\delta=-31^{\circ}38'39\arcsec$, but at this very low level, this could very
well be a false detection.}

\label{Limg}
\end{figure*}

In Fig.\,\ref{Limg} we present our \ks\ and L' band images, the nomenclature
for the clusters detected is as in C97. While all the C97 clusters are detected
in the \ks\ band, only clusters 1, 3 and 5 are clearly detected in L'.
Furthermore we find in L' a relatively bright source at about 2.4 arcsec NE of
cluster 5 which is not among the bright clusters studied in the optical by C97
and other authors. Looking in the HST images we found a faint optical
counterpart for this source. The same source is detected in our \ks\ image. We
will refer this "new" source as 5a. Very faint extended emission, possibly
resolved in 2 sources, is observable in L' to the SW of cluster 5 in 
the area of
cluster 4 that however cannot be clearly identified. In Table \ref{photom} we
present the aperture photometry of the clusters studied by C97. For the
ISAAC photometry, we did not apply aperture corrections as with the NIR PSFs,
the apertures used include $>99$\% of the flux. In addition to those 6
clusters we report the magnitudes of cluster 5a. The photometry for all
clusters has been obtained with the same method used by C97 for consistency but
with a slightly larger aperture to compensate for the larger Point Spread
Function (PSF) of our NIR observations. In the \ks\ band the galaxy background
has been measured on an annulus with radius 0.8\arcsec around each source as
done by C97 and a width of 0.7 \arcsec. In L' no other source than those
listed in Table \ref{photom} has been firmly detected, neither is any 
diffuse emission
in the galaxy. Cluster 2 and 3 are separated by less than 1 \arcsec so that
their \ks\ magnitudes are contaminated by each other.

\begin{table}

\caption{Aperture photometry obtained with a radius of 0.7\arcsec on
the main clusters detected in NGC~5253. Typical errors are less than
0.03 mag.}

\begin{tabular}{ccc}
\hline
       &  K$_{\rm s}$   &  L'   \\
\hline
  1    & 14.23 & 13.40 \\
  2    & 14.60 &   -   \\
  3    & 14.51 & 13.67  \\
  4    & 16.10 &   -   \\
  5    & 13.46 &  9.07 \\
 5a    & 15.54 & 12.51 \\
  6    & 15.64 &   -   \\
\hline
\end{tabular}
\label{photom}
\end{table}

NGC~5253 has been observed at the SEST with SIMBA on the nights of 7
and 8 of June 2002. SIMBA is a 37 channels bolometer array sensitive
in the 1.2 mm band. The beam size at the SEST is about 11\arcsec.  The
background subtraction is achieved by fast scanning of the source
through the field of view both in azimuth and altitude. Our maps had a
size of $480\arcsec \times 240\arcsec$. The atmosphere opacity $\tau$
was very good during all observations with values of about
0.1. The data have been reduced using the package MOPSI developed by
Robert Zylka. As a flux reference we have observed the planet
Uranus. The total telescope time on NGC~5253 has been about 3 hours
which gave a $1\sigma$ detection limit of 4 mJy. The SIMBA image shows
a single unresolved source whose flux is $114\pm 4$ mJy.

We complement these observations with the analysis of archival data
from the mid-IR camera of the {\em Infrared Space Observatory},
ISOCAM\footnote{The ISO archive is accessible at {\tt
http://www.iso.vilspa.esa.es/}}. The interest of ISOCAM data is
two-fold: (1) it has a much higher sensitivity than ground-based
instruments operating in the same range, and (2) with a spatial
resolution of a few arcseconds, it helps us build a more precise
spectral energy distribution for the main cluster of NGC\,5253 (see
Sect.\,\ref{sec:sed}). The data analyzed here were part of program
LMETCALF.HARO\_A and comprise the identification numbers 62500784,
62500881, and 62500980. We analyzed the observations taken with
filters LW1-[4.5\,\mic] to serve as a check on our L' observation;
LW9-[15\,\mic] to understand the discrepancies found when compiling
20\,\mic\ photometry (Sect.\,\ref{sec:sed}), and LW10-[12\,\mic] as this
filter is equivalent to the IRAS 12\,\mic\ filter and thus tells
us how much of the IRAS flux can be attributed to the central
source. All ISOCAM observations were performed with the smallest pixel
field-of-view, i.e. $1{\farcs}5$, in a ``beam-switching'' mode, i.e. a
neighboring empty field is first observed, then the telescope slews to
the source. The data were analyzed with CIA\,V5.0 following the
standard procedures. In all 3 bands the source is point-like, which
means that the transient-correction scheme used is not adapted
\citep[see][ for a discussion]{cou00}. As a result, our fluxes could
be lower limits to the actual flux, though at most by 30-50\%, given
that a very long integration is always performed on the source. The
errors quoted in Table \ref{tab:iso} are compound errors,
i.e. including photometric and transient errors, they are not
statistical 1$\sigma$. Fluxes in Table \ref{tab:iso} were obtained
with aperture photometry correcting for the missed fraction of the
PSF, as the ISOCAM PSF has more pronounced low-level wings than the ISAAC one.

\begin{table}
\caption{ISOCAM photometry of NGC\,5253}
\label{tab:iso}
\begin{tabular}{lrrrr}
\hline
Filter & FWHM ($"$) & $\lambda_{\rm ref}$ (\mic) & $\Delta\lambda$ (\mic) & Flux density (mJy) \\
\hline
LW1 & 2.3 & 4.5 & 1.0 & 125$\pm$30 \\
LW10 & 3.3 & 12.0 & 7.0 & 1960$\pm$300 \\
LW9 & 5.1 & 14.9 & 2.0 & 3900$\pm$400\\
\hline
\end{tabular}
\end{table}

\section{Colors of the Bright Clusters}
\label{sec:col}

\begin{figure}
\centering
\includegraphics[width=\textwidth]{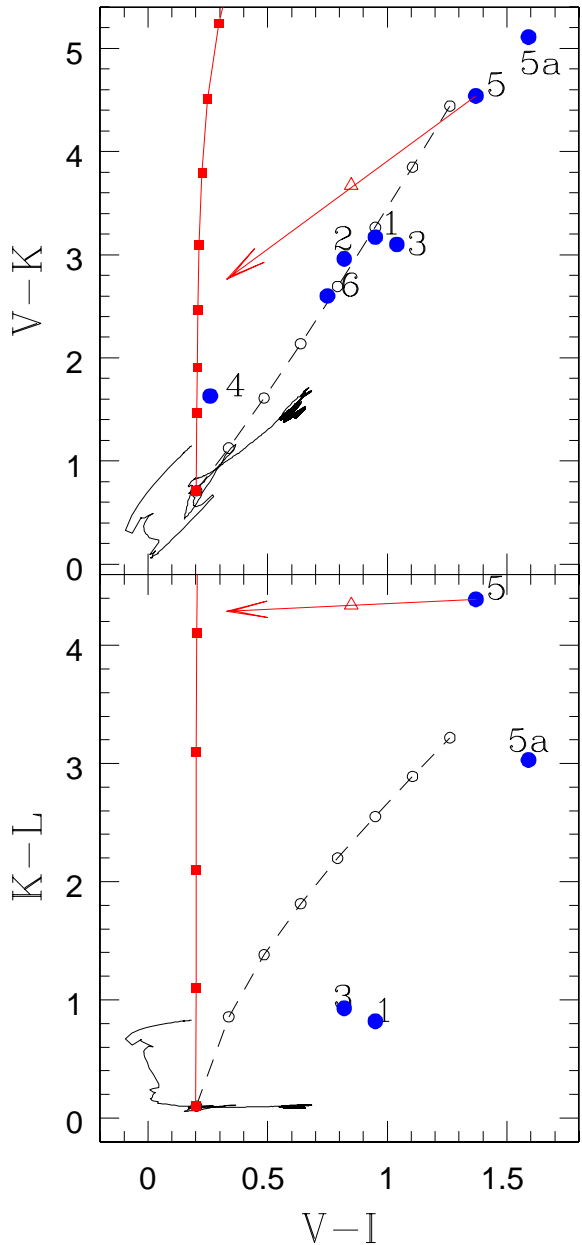}

\caption{V-K/V-I and K-L/V-I color-color diagrams of the optically
brightest clusters in NGC~5253. The solid black lines represent the
output of Starburst99 for a instantaneous burst of $Z=1/5 Z_{\odot}$. The
solid lines with square points denote the change on the 
1Myr old point introduced by an
increasing amount of dust at 1000K. The arrows indicate the effect
of 2 mag of visual extinction on the colors of cluster 5. The dashed
line with open circles shows how a more proper treatment of 
all dust effects,
extinction, scattering and emission, can affect the optical
colors. Starting from the 1\,Myr dust free Starburst99 model, we compute the
colors of the same cluster embedded in a cocoon with the dust
properties derived in Sect.~\ref{sec:sed}, with the V optical depth
increasing by 1 for each open circle, up to the cluster-5 best-fit 
value of 7.
The photometric errors on the data points are of the same size
of the symbols or smaller. 
}

\label{colors}
\end{figure}

We have combined our \ks\ and L' observations with those of C97 to
build color-color diagrams of the optically brightest clusters in
NGC~5253. For source 5a we measured the optical magnitudes on the HST
images using the same technique as in C97. To ensure
consistency between our measures we extracted the optical photometry
for all clusters from the HST images, the average difference between
our values and those of C97 is about 0.02 mag. The F547M-F814W HST
color has been converted to V-I using the standard photometric method
and a linear fit on the data provided in the HST WFPC2 photometry
cookbook\footnote{{\tt
http:/www.stsci.edu/instruments/wfpc2/Wfpcc2\_phot/wfpc2\_cookbook.html/}}. It
must be noted that while the F547M magnitudes are quite close to V,
there is a difference of more than 1 mag. between F814W and I and a
large color term. In Fig. \ref{colors} we plot the observations (solid 
circles) in
V-\ks/V-I and \ks-L'/V-I color-color diagrams.  As a reference we also
plot the output of Starburst99 \citep{lei99} - black solid line - calculated
for an instantaneous burst with a metallicity 1/5 solar and standard
IMF (2.35, 1-100 $M_{\odot}$). All observations departs significantly
from the model showing a color excess in all bands, the photometric
errors on the colors being 0.1 mag at most for the faintest sources. 
A warning note is
due at this point of the discussion. It is well known that sub-solar
single stellar population models fail in reproducing the correct blue
to red supergiants star ratio \citep{lang95} and that this has
important consequences for the predicted colors \citep{or99}. For 
this reason, interpretation of
Fig. \ref{colors} is not free from ambiguities. The problem however
only affects the objects in the supergiants phase, cluster 1 and 6
according to the ages derived by C97. In these cases, the red excess
could be artificial and partially or fully due to the lack of red
supergiants in the model.

To understand the effect that dust can have on the clusters' colors, we
examine the color corrections induced by dust emission and extinction (a third,
more complex effect, i.e. scattering, will be discussed later on). From the
colors predicted by the model for an age of 1 Myr, the square symbols
connected by a continuous line show the effect of adding to the stellar
emission an increasing fraction of emission from dust at 1000\,K (treated as a
modified blackbody with an emissivity law in 1/$\lambda$). The square
symbols indicate the fraction of Ks flux  due to dust, with  the dust
contribution doubling at each step to represent 0, 50, 67, 80, 88,
94\% etc. of the flux.  From cluster 5 we plot a screen extinction
vector corresponding to a correction for $A_{\rm V}=2$, an open triagle indicates the
value corrected for $A_{\rm V}=1$. We have used high resolution ($R=8\,10^4$) optical
spectra obtained with UVES at the VLT and still unpublished, to measure the
extinction from the Balmer decrement. Using a screen extinction model, we
have obtained $A_{\rm V}=2.65$  and $A_{\rm V}=2.95$ for cluster 1 and 5 respectively.
Dust may not present itself only as a screen but could also be mixed with the
stars, although this geometry may not be too plausible in the case of clusters.
In the mixed case, the attenuation factor is
(1-e$^{-\tau_{\lambda}}$)/$\tau_{\lambda}$. The color correction terms that
result saturate at high optical depth to values dictated by the extinction
curve that are too small to account for the observed clusters' colors. The
screen extinction vector thus represents the maximum extinction effect. 

Both color-color diagrams give consistent results. The simple examination
of Fig.\,\ref{colors} shows that the infrared excess can be explained by the
combined effect of extinction and emission by warm dust. There is a degeneracy
between age, extinction and dust fraction which prevents from accurately
defining the properties of the clusters. There are however a few important
facts: (1) an extinction of at least 1-2 mag in V affects all clusters with the
only exception of cluster 4; (2) a dust contribution in \ks\ above 30\% is
required to bring the colors of all clusters close to the model prediction
(this number is obtained by ``sliding'' the cluster points along the extinction
vector till they intersect the dust emission line); (3) clusters 5 and 5a
display the most extreme color excess: they require more than 2 mag of
extinction and a dust contribution in \ks\ above 70\%. The properties of
cluster 5 will be the main subject of the analysis presented in the following.

It must be noted that the broad band magnitudes and colors of young star
forming regions are usually strongly contaminated by the presence of emission
lines (the gas continuum emission is taken into account by the Starburst99
model). This effect is more evident in V where the contribution of the
emission lines can easily exceed 0.5 mag \citep{thu97,van03}. In our case,
however, the HST filters F547M and F814W are free from bright emission
lines so that the V-I color derived from them can be considered as
representative of the continuum. The only band that can be significantly
affected by the presence of emission lines is, in our case, \ks\ so that the
V-\ks\ colors must be considered as upper limits. Since the main emission line
present in the \ks\ band is Br$\gamma$, the effect can be quantified using the
H$\alpha$ fluxes measured by C97 and a standard value for the
H$\alpha$/Br$\gamma$ ratio. For cluster 5, even assuming an extinction 
$A_{\rm V}=9$
as derived by C97, we obtain a correction of only 0.08 mag.  The other clusters
have equivalent widths of H$\alpha$ that are considerably smaller so that the
correction for those cases can be considered negligible.  Another effect that
could potentially affect the V-\ks\ color is the different angular resolution
of the HST and VLT images which could be only partially compensated by the
different apertures used to extract the photometry.

Anticipating the results presented in Sect.\,\ref{sec:sed}, we draw the
reader's attention on the apparent disagreement between the rather small visual
extinction that can be derived from Fig.\,\ref{colors} and that derived by C97.
The reason is that the extinction vector drawn on Fig.\,\ref{colors}, as is
generally the case in such graphs, only takes into account the extinction
effect, and not the scattering by dust grains. It has long been known
\citep[e.g.][]{wit92} that while extinction reddens the intrinsic spectrum,
scattering makes it bluer. As a net result, when scattering is taken into
account, the magnitude of the extinction vector decreases for the same 
$A_{\rm V}$.
As C97 took scattering into account in their estimation of extinction, they
obtain a larger value than in the simple extinction-only case plotted in
Fig\,\ref{colors}. On the same figure, we show the color evolution of the model
derived in Sect.\,\ref{sec:sed}, with an increasing $\tau_{\rm V}$ 
(open circles, each circle marks an increase by 1 in the V optical depth, up to 
$\tau_{\rm V}=7$).
This model completely incorporates the extinction, scattering and emission
effects of dust. Because it incorporates both extinction and emission from
dust, the orientation of the color correction vector is a combination of those
of the emission vector (squares) and screen extinction (triangle) vector. But
since it also accounts for scattering, the amplitude of the color correction
for one mag. of optical depth is smaller than in the pure screen extinction
case. As will be seen in
Sect.~\ref{sec:sed}, our model only partially reproduces the L' band flux,
hence the disagreement in the \ks-L' color.

All clusters are barely resolved in our images. In particular in \ks\, where
all field stars display a uniform PSF with FWHM of $0\farcs40$, clusters 1 and
6 have a FWHM of $0\farcs50$, cluster 5 of $0\farcs46$, cluster 3 of
0$\farcs$52, and cluster 2 of 0$\farcs$60, though this last value is made
uncertain by the proximity of cluster 3. Cluster 4 is too faint to measure a
meaningfull FWHM. Deconvolving with the stellar PSF we obtain sizes in the
range $0\farcs22-0\farcs45$ (3.5-7.2pc) that are fully consistent with
those measured by C97. In the L' image there are no stars visible so that we
have to rely on the standard stars which have a FWHM of $0\farcs4$. Cluster 5,
the only one for which we can reliably measure the FWHM in L', has a FWHM of
$0\farcs47$ consistent with the value measured in Ks. This means that the dust
emission is very compact in \ks\ and L' and that it is confined to the
optically emitting region.

\section{Modeling the SED}
\label{sec:sed}

As it is evident from figure\,\ref{colors}, cluster\,5 stands out among
the population of bright stellar clusters in NGC\,5253. Since cluster\,5
sits at the position of the thermal radio nebula of \citet{tur00}, we
can assume that it is indeed the optical counterpart of the radio
nebula. The fact that optical studies show it to suffer from strong
extinction strengthens this hypothesis. Thus in this section we first
compile observations to build the full spectral energy distribution of
cluster\,5, from the optical to the radio, and then we attempt to
model this SED using the same approach as in SBS\,0335-052
\citep{pla02}, i.e. modeling the stellar light transfer in the dust
envelope with DUSTY \citep{ive97}.

\subsection{The SED of cluster 5}

From the optical to the L' band our task is simple since the spatial resolution
of HST (C97) or VLT (this work) allows to separate the cluster from its
neighbors and to measure accurately its fluxes. Our optical
photometry of cluster 5 agrees within 0.06 mag with C97, the difference
is mostly due to the background subtraction.

In the mid-IR range, fluxes come from a variety of works, from the early ones
with photometers of \citet{rie72}, \citet{moo82}, and \citet{fro82}, to
state-of-the-art cameras on large ground-based telescopes \citep{gor01} and
space observatories, namely IRAS and ISO. At the shortest infrared wavelengths,
the ISOCAM 4.5\,\mic\ flux of 125$\pm$30\,mJy compares well with the 77\,mJy
measured in the L'(3.78\,\mic) image, once taken into account the strong rise
from \ks\ to L'. There is thus no reason to doubt that all of the ISOCAM
4.5\,\mic\ flux comes from cluster\,5. Around 10\,\mic\ the measurement from
\citet{rie72} appears rather high compared with the more recent determinations
of both \citet{fro82} and \citet{moo82} and we thus choose to exclude it from
the SED. As the SED is seen to increase as the wavelength increases, the
2.2\,Jy determination at 11.7\,\mic\ of \citet{gor01} can be considered
compatible with the rest of the photometry, although on the high side. The
comparison of our LW10 and the IRAS 12\,\mic\ flux shows that the ISO flux
represents only $\sim$75\% of the IRAS flux. Given that dust is clearly seen to
be widespread in the galaxy, we attribute this discrepancy
to diffuse dust, to which ISO and ground based telescopes are much less
sensitive than IRAS. Diffuse dust, also called cirrus dust in the Galactic
context, has its emission peak around 100\,\mic. Therefore it must contribute
emission to all four IRAS bands, meaning that we cannot attribute 100\% of the
IRAS fluxes at 25, 60 and 100\,\mic\ to cluster\,5. As we do not have high
spatial resolution measurements at these wavelengths, we assume that the same
correction factor applies to them and consider that only 75\% of the IRAS flux
at 25, 60 and 100\,\mic\ originates from cluster\,5.

Moving to the range 15-25\,\mic, the situation becomes slightly
confused. As the 18\,\mic\ range is visible from the ground, a number
of measurements exist, but they show a rather large scatter, from the
small value of 2.9\,Jy obtained at 18.7\,\mic\ by \citet{gor01} to the
large IRAS 25\,\mic\ flux of 9\,Jy (75\% of 12\,Jy). Even though the
wavelengths are not the same, an increase by more than a factor of 4
is quite surprising. In fact we suspect that, although it is backed
by some earlier measurements, the Gorjian et al. flux underestimates
the actual flux in this range. For instance, the point source detected
with ISOCAM at 14.9\,\mic\ has a flux of 3.9$\pm$0.4\,Jy, already
higher than the 18.7\,\mic\ flux, and very few sources show such a
decreasing spectrum from 14.9 to 18.7\,\mic, even when taking into
account the second silicate absorption band at 18\,\mic. Furthermore,
and although the photometric accuracy on the continuum is probably not
very good, the spectrum of \citet{cro99} shows that the continuum
increases slightly from 15 to 19\,\mic, with the level at 25\,\mic\
being a good factor of 2 higher than at 18\,\mic. The low Q band flux
of \citet{rie72} at 3.7\,Jy could lend some support to the
\citet{gor01} measurement however \citet{rie72} give no detail on the
band used for this measurement and thus we cannot use it to constrain
the SED. According to Gorjian (private communication) the low
18.7\,\mic\ flux could be due to too small a chopping throw, resulting
in an incorrect zero-level, or a complex airmass correction. We note
that a similar measurement in the 18.7\,\mic\ filter was obtained by
\cite{mar04} giving a higher flux of 5.7\,Jy. We will thus discard the
18.7\,\mic\ measurement of \citet{gor01} from the SED fit.

Further on in the infrared, we only have IRAS measurements, for which
the PSF is almost as large as the galaxy. Again we will attribute only
75\% of the IRAS flux at 60 and 100\,\mic\ to cluster\,5.

Coming to the submillimeter and millimeter range, we enter a realm where dust
continuum emission is not necessarily the dominant emission process. Indeed
synchrotron or free-free emission usually dominate longward of a few cm. Here
we will follow the computations of \citet{tur97}. These authors show that
longward of 2\,cm, and for the compact radio source, the flux can reasonably be
considered as pure free-free emission. This process has a spectral index of
$\alpha=-0.1$ (in $f_{\nu}\propto\nu^{\alpha}$). From the 2\,cm flux they
compute the remaining dust continuum emission at 2.6\,mm, an upper limit of
26$\pm$10\,mJy. This will be the longest wavelength included in the SED, and
given that it is based on interferometric data we attribute all this flux to
cluster\,5. Using the same method we estimate the free-free contribution in our
1.2\,mm observation and in the \citet{jam02} 850\,\mic\ SCUBA observation at
respectively 43 and 41\,mJy. For those last two measurements however we are
left in a situation intermediate between the optical and the IRAS ones: at
$\sim$10-20\arcsec, the PSF could be rather large compared to the object.
Judging from the \citet{jam02} map, the SCUBA emission comes from the central
region of the galaxy, where cluster\,5 is located, but also where most of the
star formation activity, and ionized gas are concentrated \citep[as revealed
e.g. by the radio and H$\alpha$ images of][]{tur98}. This is also the case in
our 1.2\,mm observations. The association with the ionized gas could simply
reflect the free-free contribution ($\sim$ 20\%) but we have no means of
ascertaining that. Therefore we will attribute to cluster\,5 the full
free-free-corrected fluxes at 1.2\,mm and 850\,\mic, i.e. respectively 71$\pm$4
and 151$\pm$23\,mJy. The errors quoted here are the formal errors derived
from ours or \citet{jam02} photometry. A further source of error comes from the
free-free correction, however \citet{tur97} do not provide enough details for
us to estimate the associated error.

\begin{figure*}

\includegraphics[angle=0,width=\textwidth]{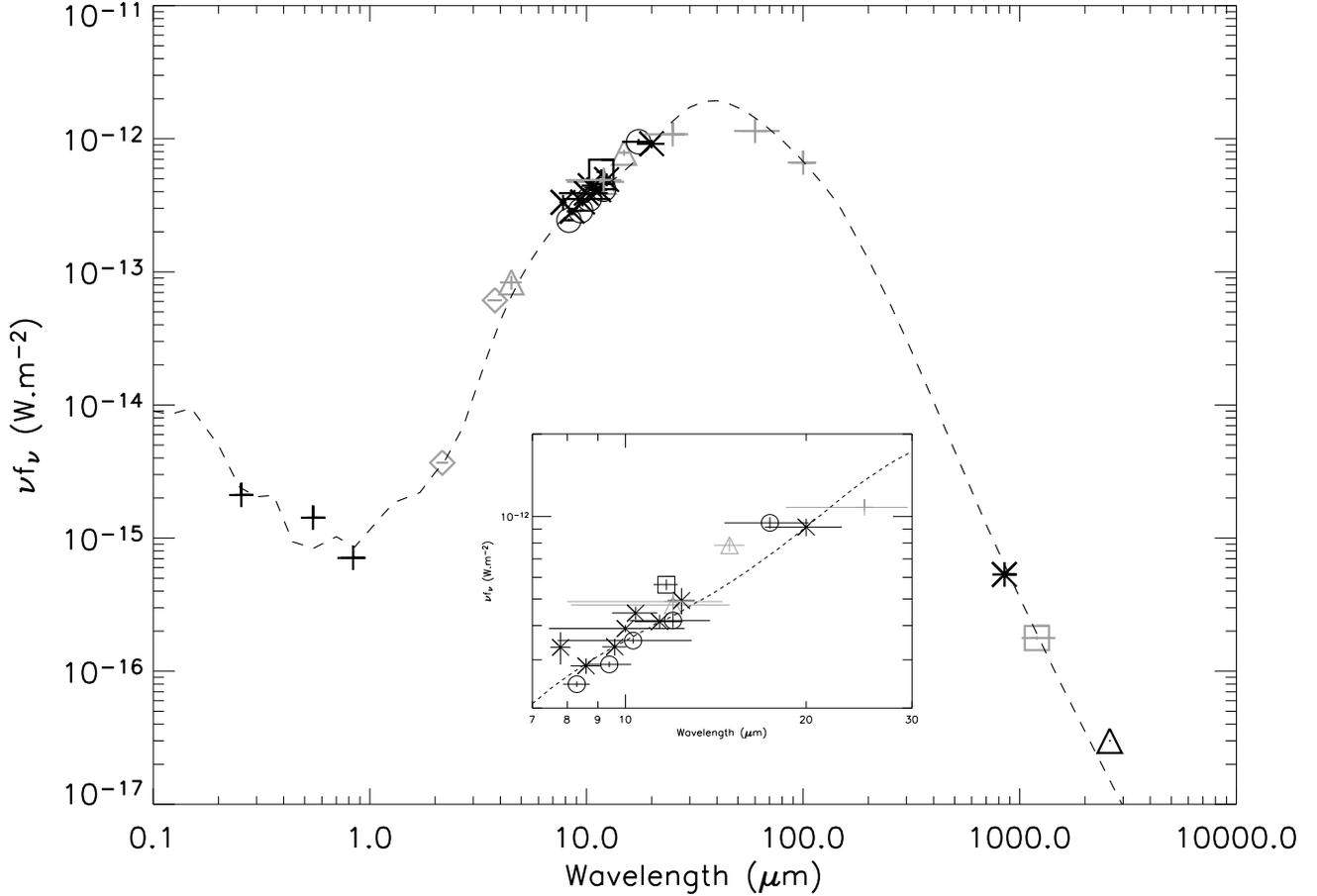}

\caption{The spectral energy distribution of cluster\,5 in
NGC\,5253. The 7-30\,\mic\ range has been blown up for clarity in the inset. 
Symbols refer to the photometric source, 
from left to right: black plus signs correspond to the HST 
photometry, grey diamonds to the VLT/ISAAC data, grey triangles to the 
ISO data, black crosses in the 10\,\mic\ range to the \citet{fro82} 
data, black circles to the \citet{moo82} data, the black square to the 
\citet{gor01} 11.7\.\mic\ point, grey plus signs to IRAS data, the 
black cross in the submillimeter range to SCUBA data, the grey square 
to our SEST/SIMBA observation, and the black triangle is the OVRO upper
limit. Horizontal error bars describe the filter bandwidths, while
vertical error bars show the photometric errors, as quoted by the
respective authors (see Table\,\ref{tab:sed}). In most cases, these
bars are shorter than the symbol size. The dashed line is the best
fitting DUSTY model.}

\label{fig:sed}
\end{figure*}

Table\,\ref{tab:sed} and Fig.\,\ref{fig:sed} summarize all the
measurements included in the spectral energy distribution of
cluster\,5.

Judging from Fig.\,\ref{Limg} one could think that the existence of
source 5a brings a further uncertainty to the SED we have just
built. We do not think that this is the case since already in L' the
source is 3.5\,mag fainter, and is less red than
cluster\,5. Furthermore, it does not appear to be detected with
ground-based telescopes at 10\,\mic.

\begin{table*}
\caption{The compiled spectral energy distribution of
Cluster\,5. Aperture size are listed when appropriate (i.e. mostly for
single-beam measurements). A few telescopes and filter bandwidths are
missing because we were unable to find the information in the original
papers.}
\label{tab:sed}
\begin{minipage}{18cm}
\begin{tabular}{llrrrrll}
\hline
Telescope/instrument & Filter name & $\lambda$ & $\Delta\lambda$ & aperture 
& Flux & Source \\
 & & \mic & \mic & \arcsec & mJy & \\
\hline
HST/WFPC2      & F255W & 0.2574 & 0.041 & & 0.179 & C97 \\
HST/WFPC2      & F547M & 0.5465 & 0.049 & & 0.259 & C97 \\
HST/WFPC2      & F814W & 0.8370 & 0.176 & & 0.198 & C97 \\
VLT/ISAAC      & Ks    & 2.16   & 0.27  & & 2.65 & this work \\
VLT/ISAAC      & L'    & 3.78   & 0.58  & & 77.2 & this work \\
ISO/ISOCAM     & LW1   & 4.5    & 1.0   & & 125$\pm$30 & this work \\
CTIO 1.5 \& 4m &       & 7.8    & 0.6   & 8.2 & 866$\pm$234 & \citet{fro82} \\
ESO 3.6m       & N1    & 8.3    & 0.85  & 7.5 & 676$\pm$32 & \citet{moo82} \\
CTIO 1.5 \& 4m &       & 8.6    & 1.0   & 8.2 & 817$\pm$106 & \citet{fro82} \\
ESO 3.6m       & N2    & 9.4    & 1.65  & 7.5 & 904$\pm$43 & \citet{moo82} \\
CTIO 1.5 \& 4m &       & 9.6    & 0.9   & 8.2 & 1070$\pm$150 & \citet{fro82} \\
CTIO 1.5 \& 4m &       & 10.0   & 5.1   & 8.2 & 1300$\pm$80 & \citet{fro82} \\
               &       & 10.0   & 5.4   & 6.0 & 2130$\pm$150\footnote{This 
measurement is not used as a constraint for the SED fit, see text for 
details.} & \citet{rie72} \\
ESO 3.6m       & N     & 10.3   & 5.2   & 7.5 & 1210$\pm$60 & \citet{moo82} \\
CTIO 1.5 \& 4m &       & 10.4   & 1.8   & 8.2 & 1540$\pm$170 & \citet{fro82} \\
               &       & 10.4   & 1.3   & 8.5 & 1365$\pm$90 & \citet{leb79} \\
               &       & 10.6   & 5     & 8.5 & 1500 & \citet{leb79} \\
CTIO 1.5 \& 4m &       & 11.4   & 2.1   & 8.2 & 1570$\pm$190 & \citet{fro82} \\
Keck/LWS       &       & 11.7   & 1.0   & & 2200$\pm$200 & \citet{gor01} \\
ESO 3.6m       & N3    & 12.0   & 3.7   & 7.5 & 1670$\pm$200 & \citet{moo82} \\
ISO/ISOCAM     & LW10  & 12.0   & 7.0   & & 1900$\pm$300 & this work \\
IRAS           & 12    & 12.0   & 7.0   & & 2610$\pm$131\footnote{Only 75\% 
of this flux is used as a constraint for the SED fit, see text for details.} 
& \citet{mos90} \\
CTIO 1.5 \& 4m &       & 12.4   & 1.3   & 8.2 & 2040$\pm$450 & \citet{fro82} \\
ISO/ISOCAM     & LW9   & 14.9   & 2.0   & & 3900$\pm$400 & this work \\
ESO 3.6m       & Q     & 17.4   & 5.6   & 7.5 & 5500$\pm$539 & \citet{moo82} \\
Keck/LWS       &       & 18.7   & 0.5   & & 2900$\pm$300$^{a}$ & 
\citet{gor01} \\
CTIO 1.5 \& 4m &       & 20.0   & 5.9   & 8.2 & 6100$\pm$900 & \citet{fro82} \\
               & Q     & 21.0   &       & 6.0 & 3700$\pm$400$^{\rm a}$ & 
\citet{rie72} \\
IRAS           & 25    & 25.0   & 11.15 & & 12000$\pm$600$^{\rm b}$ & 
\citet{mos90} \\
IRAS           & 60    & 60.0   & 32.5  & & 30500$\pm$1220$^{\rm b}$ & 
\citet{mos90} \\
IRAS           & 100   & 100.0  & 31.5  & & 29400$\pm$1760$^{\rm b}$ & 
\citet{mos90} \\
JCMT/SCUBA     &       & 850.0  & 62    & 41 & 151$\pm$23\footnote{Corresponds
to the dust-only flux, see text for details.} & \citet{jam02} \\  
SEST/SIMBA     &       & 1200.0 & 465   & & 71$\pm$4$^{\rm c}$ & this work \\
OVRO           &       & 2600.0 & 22.6  & & $<26^{\rm c}$ & \citet{tur97} \\
\hline
\end{tabular}
\end{minipage}
\end{table*}

\subsection{Fitting the SED}
\label{sec:fitsed}

Now that the spectral energy distribution has been built, we turn to
DUSTY \citep{ive97} to model it. The main feature of DUSTY is that it
models the radiative transfer of light emitted from a point source
inside a spherically symmetric shell of dust. Though straightforward
to describe, this problem actually involves a relatively large number
of parameters. The first one is obviously the light source. Here,
following Sect.\,\ref{sec:col}, we will use
Starburst99 \citep{lei99} to generate instantaneous burst models with
ages from 1 to 9\,Myr at a metallicity of 1/5 $Z_{\odot}$, close 
to the one of NGC~5253. Describing the dust shell requires many more
parameters. The first of these is the dust equilibrium temperature at
the inner side of the shell, $T_{\rm in}$. Due to the strategy developed
by DUSTY to solve the transfer problem, this temperature sets the
physical scale of the model. The radial profile of the dust density
has to be described. We take it as a broken power law. For $N_{\rm rad}$
radial zones, this generates $(N_{\rm rad}-1$) radius values for the zone
limits as the outer radius is fixed in terms of the inner radius, and
$N_{\rm rad}$ exponents. To introduce some flexibility in the density
profile with a limited number of parameters, we chose to run models
with two radial zones. In principle, DUSTY can include six different
dust grains species. However these are mainly variations around three
principal types: the classical astronomical silicates and graphite
\citep{dra84}, and amorphous carbon grains, a possibly more realistic
representation of interstellar carbon-based dust. With three grains
species, the dust composition introduces 2 parameters. Dust is also
characterized by its size distribution, taken as a power-law, and thus
described with three parameters, the minimum and maximum grain size,
and the exponent of the power law. The standard size distribution, the
so-called MRN distribution has $a_{\rm min}=0.005\,\mic$,
$a_{\rm max}=0.25\,\mic$ (1\,\mic\ for graphite) and an exponent $q=3.5$
\citep[in the sense $n(a)\propto a^{-q}$,][]{mat77}. This size
distribution is such that, for a 50-50 mix of astronomical silicate
and graphite, it reproduces the Galactic interstellar extinction
curve. Finally, the last parameter describing the shell is its
thickness, characterized in DUSTY by the visible optical depth
$\tau_{\rm V}$.

Therefore to summarize the parameters that describe a DUSTY model in
our case are: $t_{\rm sb}$ the age of the burst, $T_{\rm in}$ the temperature
on the inner side of the shell, $X_{\rm Si}$ and $X_{\rm gr}$ the abundances
of silicates and graphite (the abundance of amorphous carbon is $X_{\rm am}=1-[X_{\rm Si}+X_{\rm gr}]$), $a_{\rm min}$, $a_{\rm max}$, and $q$ the
parameters of the size distribution, $r_{\rm lim}$ the transition radius
between the two radial zones, $\beta_{1}$ and $\beta_{2}$ the
exponents of the density power-law (in the sense $\rho \propto
r^{-\beta}$), and finally $\tau_{\rm V}$ the optical depth. This is a
total of 11 parameters. As can be seen in Table\,\ref{tab:sed} we have
almost 3 times as many photometric data points so the problem is
clearly over-constrained. However an 11-dimension parameter space is
not one that is easy to explore systematically. To identify the
best-fit model, we choose a strategy where we start from a standard
dust shell model and have a restricted set of parameters to explore their
full space. The best fitting model among this set is then used as a
``seed'' and a new set of parameters is allowed to explore a large
part of their respective space. In general, when we start with a new
seed model, parameters of the previous exploration are still allowed
some margin. This is especially true of $\tau_{\rm v}$ for purely technical
reasons. This is repeated a number of times so that all the 11
dimensions of the parameter space are explored more than once.

To compare a model to the observations, the model SED is integrated
using the filters' respective transmission curves and photometric
conventions to produce model fluxes. For the photometric data of
\citet{moo82} and \citet{fro82}, we could not find the actual filter
transmissions and thus a square bandpass was used. The best fitting
model is the one that minimizes the quantity:

\begin{equation}
\Delta = \sum_{i=1}^{N_{\rm filter}} \left[ \frac{f_{i}^{\rm obs}-f_{i}^{\rm mod}}
{f_{i}^{\rm obs}} \right]^2
\label{eq:diff}
\end{equation}

The advantage of Eq.\,(\ref{eq:diff}) is that it gives an equal weight
to the small and large flux values, an important fact given the very
large dynamic range of cluster\,5's SED (about 6 orders of 
magnitude), and that it is symmetric, i.e. models that over- or 
under-predict the SED by the same amount have equal weight.

The best-fitting model is displayed as a dashed line on
Figure\,\ref{fig:sed}. The search for a better fit in an 11-dimension
parameter space can be a never-ending one. This model is the
best-fitting in the sense that it reproduces the data within the average 
photometric error, with the exception of the F547M filter. In
Table\,\ref{tab:param} we have indicated the range in which individual
parameters can vary without significantly degrading the fit.  Before
extracting physical information from the fit, it is worth commenting
on some of the features discovered as we searched for the best fitting
model.

First, while we have allowed the age to vary between 1 and 9\,Myr for
a substantial part of the parameter space exploration, we have found
no intermediate best-fitting model (or ``seed'' model to use the
terminology above) for ages different than 1 or 2\,Myr. This age is
younger than that derived from the optical colors alone, although not
by a large amount. It is also smaller than the age
of the Antennae dust-enshrouded SSC \citep{gil00}, making it one of
the youngest such source.

Second, the optical depth converges invariably toward a value of
7-8. Interestingly, this is very close to the value measured by C97,
9\,Vmag. It is also in the same range as that measured by \citet{gil00} 
for the Antennae SSC. However it does not agree with what
can be deduced from the color-color diagram presented in
Figure\,\ref{colors}: the red arrow extending from cluster\,5 seems to
imply only 2\,Vmag of extinction. The reason is simple: the effect of
dust on the flux in the V, I, \ks\ and L' bands is much more complex
than a simple screen extinction on the stellar flux combined with
thermal emission in the infrared bands. Scattering also has to be
taken into account and, as it is amply demonstrated in \citet{wit92},
it makes the intrinsic spectrum bluer, and thus works in the
opposite direction of extinction. This very complex effect of dust is
exemplified on Fig.~\ref{colors} with the dashed line: we have taken
the best fitting model and decreased its visual optical depth
($\tau_{\rm V}$) from 7 to almost 0 (i.e. no dust at all). As can be seen,
the effect on the colors combines that of the simple screen extinction
and hot dust contamination, but the color change per unit of optical
depth is much less than that implied by screen extinction, mostly
because of scattering.

Finally the temperature on the inner side of the shell is also rather
stable in all seed models at around 500-600\,K, with the best-fit
value at 570\,K. When compared with the results obtained for
SBS\,0335-052 \citep{pla02}, i.e. an optical depth of 30 and an
internal temperature of 700\,K, this shows that the cluster in
NGC\,5253 is a less extreme case than that of SBS\,0335-052, though
not strikingly different.

\begin{table}
\caption{Parameters of the best-fitting dusty model. In the acceptable
range, we give an indication of the range of values that, while
producing a worse fit, still correspond to a model reproducing the
data within the photometric uncertainties.}
\label{tab:param}
\begin{tabular}{lll}
\hline
Parameter & Value & Acceptable Range \\
\hline
$t_{\rm sb}$    & 1\,Myr             & 1-2 Myr \\
$T_{\rm in}$    & 570\,K             & 550-650\,K \\
$X_{\rm Si}$    & 0                  & $<10$ \\
$X_{\rm Gr}$    & 0.84               & 0.8-0.9 \\
$a_{\rm min}$   & 0.005\,\mic\       & $<0.01$\,\mic \\
$a_{\rm max}$   & 0.5\,\mic\         & 0.4-0.6\,\mic\ \\
$q$         & $2.5$              & $2.5$-$3.0$ \\
$r_{\rm lim}$   & 2.8\,pc            & 2.8-7\,pc \\
$\beta_{1}$ & 0                  & 0-0.5 \\
$\beta_{2}$ & 0.5                & 0-0.5 \\
$\tau_{\rm V}$  & 7                  & 7-8 \\
\hline
\end{tabular}
\end{table}

From the actual parameters of the best-fitting model (see
Table\,\ref{tab:param}), we can extract physical parameters of the
embedded source and its surrounding cocoon \citep[the equations to
perform this can be found in][]{pla02}. The best-constrained parameter is
the luminosity since this is simply the integral of the SED, which is
already well described by the observations. The luminosity we obtain
is 1.2\,$10^{9}$\,L$_{\odot}$. The uncertainty on this value is
$\sim$20\%. The bolometric luminosity of a nearby galaxy is a very
difficult quantity to estimate, however one can get a good insight of
the importance of cluster\,5 by comparing its SED,
i.e. Figure\,\ref{fig:sed}, to the complete SED of the galaxy, readily
available on the NASA Extragalactic Database\footnote{see {\tt
http://nedwww.ipac.caltech.edu/}}. This shows, quite expectedly, that
cluster\,5 produces nearly all of the IR-Submm luminosity of
NGC\,5253, but more important, it produces a luminosity equivalent to
that emerging in the UV-visible region. Therefore this shows that a
single source, of very modest physical size, can generate as much
energy as the complete galaxy. In other terms, it means that even when
we consider the global SED of the galaxy, the origin of the energy can
be a very strong function of the wavelength, i.e. from the widespread
distribution of stars over the whole galaxy body in the UV-visible, to
a single compact source in the IR-Submm. Depending on the
representativity of NGC\,5253 with respect to other galaxies, this
fact could have a significant impact on all attempts to model the
evolution of global SED \citep[e.g. ][]{cha01,dal02,cha03}. Finally,
to compare clusters to one another, it is often convenient to express
the luminosity as a number of equivalent O7V stars \citep[see the
properties of this ``standard'' in][]{vac96}. We find that the
bolometric luminosity of cluster\,5 requires 4700 of these stars, well
in the range of SSC's stellar content and quite close to the radio
estimate of 4000 \citep{tur00}.

Using this luminosity and Starburst99 we can obtain the mass of stars present
in the cluster. This is 8.2\,10$^{5}$\,M$_{\odot}$, very close to the estimate
of C97 (10$^{6}$\,M$_{\odot}$). This mass obviously depends on the IMF used in
the model, in this case a Salpeter IMF with masses between 1 and
100\,M$_{\odot}$, and would increase to 2.1\,10$^{6}$\,M$_{\odot}$ if the mass
distribution extends down to 0.1\,M$_{\odot}$. However, assuming a single
slope value over the whole 0.1-100\,M$_{\odot}$ range is probably not justified
\citep[see for instance the detailed study of][]{kro93} and usually
overestimates the stellar mass. Using a slope 1.25 between 0.1 and 1
\,M$_{\odot}$ \citep{scalo86}, the mass of the cluster becomes
1.2\,10$^{6}$\,M$_{\odot}$. This mass is well in the range observed for SSCs,
although compared to SBS\,0335-052 or the Antennae, it appears on the low side
\citep[see][]{pla02,gil00}.

Concerning the geometry of the cocoon around the cluster, we find that
it extends out to a radius of $\sim$140\,pc, or approximately
9$\arcsec$. This value strongly depends on $T_{\rm in}$ and it is smaller
for larger values of the inner temperature.  We point out that this is the
predicted size of the dust cloud inside which the cluster should
reside, but that depending on the observing wavelength, the angular
size of the source will vary, since the dust temperature strongly
decreases with radius. Even though, this is rather large compared to
the size of the central star formation region of the galaxy (see
Fig.~\ref{Limg}).  This can be resolved in a number of ways. (1)
Cluster\,5 is only located in projection in the central region, but it
is in fact behind, or less likely in front of the star forming region,
or (2) cluster\,5 is indeed part of the star forming region, and the
dust cloud we model here encompasses it all. In that second case,
which is more likely, all the central clusters should suffer from some
extinction, which is the case, and all these provide some heating to
the dust, so that our SED for cluster\,5 is overestimated. Correcting
this is impossible, but would lead to a lower luminosity for the
central cluster and a smaller stellar mass. We note though, that the
high resolution observations in the 10-20\,\mic\ region show that only
cluster\,5 contributes infrared emission. This means that the heating
provided by the other clusters is likely small, hence the correction
on the model result should be small as well.  It is interesting to
note that the predicted size should be close to the observed size in
the submillimeter range, since in that range we are most sensitive to
cold dust. Indeed the SCUBA map of \citet{jam02} shows a source with a
compatible angular size (25-30$\arcsec$ diameter at the 6$\sigma$ level).

The most striking properties of the dust cocoon, as found from the best-fitting
model, are directly connected to the dust grains. We find that the exponent of
the grain size distribution is $2.5$, shallower than in the MRN distribution,
i.e. it biases the distribution toward larger grains compared to the standard
interstellar medium, as was found for \sbs. As noted in \citet{pla02} this 
appears to be common in high density regions exposed to strong radiation
fields \citep[see for instance the work of][]{mai01}. With sizes between
0.005\,\mic\ and 0.5\,\mic, we do not find in NGC\,5253 the lack of small
grains that was observed in \sbs\ where sizes ranged between 0.2 and 1\,\mic.
Finally the chemical composition of the dust is very different from that found
in the general ISM of our galaxy: we find that for the best-fit model, the
dust composition requires no silicate component, being made of 84\% graphite
and 16\% amorphous carbon. This is an extreme composition, even more severe
than in the \sbs\ cases where the best fit model required 23\% silicates
\citep{pla02}. It first finds some observational support in the SWS spectrum of
\citet{cro99}: even though independent measurements show significant extinction
toward cluster\,5, the spectrum does not show any absorption at the location of
the strongest silicate band, 9.7\,\mic. We have then explored the deterioration
of the fit if we intentionnaly add silicates to the dust mixture, while keeping
the rest of the model constant. We observe that the figure of merit of the fit
degrades gradually, being twice as large as the best fit value for a 25\%
abundance of silicates. However the worst problem is that since our model
requires hot grains toward the center of the structure to fit the NIR-MIR
emission, as soon as the fraction of silicates reaches 10\%, we start to
observe emission features in the model, which are absent from the SWS spectrum
for instance. We therefore feel that indeed the dust in the cocoon presents a
severe lack of silicates.

We find that the density profile along the shell radius is shallower
than in \sbs. Here the shell has a flat profile over the first 2-5\%
of its radius and then falls with an exponent of $1/2$. This profile,
combined with the dust composition and size distribution, leads to a
total dust mass in the envelope of 1.5\,10$^{5}$\,M$_{\odot}$. The
dust mass is directly proportional to the optical depth. It obviously
also depends on the chemical composition, dust size distribution and
density profile, all three parameters also influencing the optical
depth. We find that the acceptable range of values can lead to a
variation of the dust mass by a factor of $\sim$2.  \citet{mei02}
estimate that $\sim10^{7}$\,M$_{\odot}$ of H$_{\rm 2}$ is present in
the central region of the galaxy. Not all this gas is associated to
the star formation episode: if we follow the estimation of the same
authors, only 5\,10$^{5}$\,M$_{\odot}$ of H$_{\rm 2}$ are causally
connected to the current starburst, which obviously includes
cluster\,5. This would imply a molecular gas to dust mass ratio of
$\sim$3-4. This is rather high, even considering the lack good
statistics on this particular ratio. It is however not possible to
elaborate on this particular issue, given the facts that (1) the
molecular mass comes from an interferometric measurement, (2)
NGC\,5253 is a low-metallicity star-forming galaxy where the
CO-to-H$_2$ conversion ratio can be quite different from the standard
value and (3) the determination of which fraction of the H$_{\rm 2}$
gas mass is actually linked with the current burst is a difficult
matter.  We also note here that the mass of dust would be decreased if
DUSTY allowed for temperature fluctuations in the dust
\citep[see][]{pla02}. If we consider that the dust mass in the cocoon
is a good measure of the total dust mass in the galaxy (the cocoon
provides almost all the IR-Submm luminosity), then the molecular gas
to dust mass ratio in the central region of the galaxy becomes
$\sim$70, i.e. a more reasonable value.

Finally, we note that in a recent investigation of the ionized nebula
associated with cluster~5, \citet{tur03} consider that none of the optical
clusters, those detected in the HST data, coincide with the near-infrared
source. This rests on a displacement between cluster~5 in the HST data and the
NIR main source of $0\farcs3\pm0\farcs1$, which we also confirm with our data.
Does this mean that indeed the two objects, cluster~5, and the infrared source
are different? We think this is too small an offset to guarantee that indeed we
have two independent sources here. The visual location of the source could for
instance be affected by non-uniform extinction on the face of the cluster. To
be complete however we investigated the consequences \citet{tur03} statement,
which basically result in the HST data dropping out of the fit. The main
consequence is that the optical depth has to increase to values in the range
15-20, which in turn increases the dust mass required in the enveloppe. We
therefore conclude that the model we have presented here represents at least a
lower limit (in terms of the mass involved) to the actual situation in
NGC\,5253. To be able to state whether we are dealing with a single source from
the visible to the NIR or not would require high-resolution images covering the
V to K band, in order to ``follow'' the source's morphology and position with
wavelength.

\section{Conclusions}
\label{sec:conc}

We can summarize the conclusions of our work as follows:

\begin{itemize} 

\item We have obtained very high quality NIR data of NGC\,5253 in the
\ks\ and L' bands. The optically bright clusters detected all show a red
excess that can be explained with the combined effect of extinction
and dust emission. This however must be considered cautiously for the
clusters in the red supergiant phase due to the uncertainty introduced
on the models by the metallicity.

\item One single cluster dominates the galaxy output beyond
3\,\mic. We have detected NGC\,5253 at 1.2 mm, we attribute the flux
observed at this wavelength to the IR bright cluster known as
cluster 5.

\item Using our observations, archive data and data available in the
literature we have built a complete spectral energy distribution from
the optical to the millimeter, doing this we find further evidence that
one single cluster is responsible for most of the IR flux which is
comparable to the optical flux of the whole galaxy.

\end{itemize}

We have then used the model DUSTY to fit the SED and derive the
parameters of the IR cluster. We have found that:

\begin{itemize}

\item The IR cluster has a stellar mass of about 
0.82-2.1\,10$^{6}$\,M$_{\odot}$ depending on the IMF used, and a
luminosity equivalent to 4700 O7V stars. This cluster is obscured by 7 mag of
optical extinction.

\item The dust in which the cluster is embedded shows the now familiar bias
toward shallower size distribution than the MRN one, and strikingly lacks
silicates, a composition that can explain the absence of a significant silicate
absorption band. Adding silicates to the best fit model, to  obtain a dust
composition that resembles the classical one, lead to emission features in the
model spectrum as the NIR part of the SED requires relatively hot grains. None
of these features are observed.

\item The total dust mass necessary to reproduce the optical to submm SED of
cluster\,5 is 1.5\,10$^{5}$\,M$_{\odot}$. This is a relatively small amount
intrinsically, although when compared to the molecular mass observed to be
involved in the current episode of star formation, it becomes a substantial
fraction of the ISM.

\end{itemize}

Finally we wish to stress again that NGC\,5253 provides a striking example of
an object where one part of the spectral energy distribution, i.e. the
IR-submm one, is completely decoupled from the other, the UV-optical one. The
existence of such objects, if not a space oddity, would cast some doubt on our
ability to predict the global spectral energy distributions of galaxies as they
evolve in time.

\begin{acknowledgements} 

We are grateful to Daniela Calzetti, Claus Leitherer and St\'ephanie
Plante for providing us useful information during our work. We thank
Daniel Schaerer and Leticia Martin-Hern\'andez for letting us quote 
their new NGC\,5253's Q1-band flux. We also thank
thanks the ESO staff that carried out the observations at the VLT in
Service Mode providing us data of outstanding quality. MS acknowledges
the support of ESO under the Visiting Scientist program.
This research has made use of the
NASA/IPAC Extragalactic Database (NED) which is operated by the Jet
Propulsion Laboratory, California Institute of Technology, under
contract with the National Aeronautics and Space Administration. The
ISO data used in this paper was processed with CIA, a joint
development by the ESA Astrophysics Division and the ISOCAM consortium
led by the ISOCAM P.I. C. Cesarsky, Direction des Sciences de la
Mati\`ere, C.E.A. Saclay, France. Finally we thank the referee,
Richard de Grijs, for his comments which contributed to improve this 
paper and for his very positive attitude and availaibility to discussion.

\end{acknowledgements}

\end{document}